\documentclass[prc,aps,twocolumn,nofootinbib,superscriptaddress,showpacs]{revtex4-1}
\usepackage{graphicx}
\usepackage{amssymb}
\usepackage{amsmath}
\usepackage{dcolumn}
\usepackage{bm}
\begin{document}

\title{Neutrino-nucleon scattering in supernova matter from the virial expansion}

\author{C.\ J.\ Horowitz}
\email[E-mail:~]{horowit@indiana.edu}
\affiliation{Center for Exploration of Energy and Matter and Department of Physics, Indiana University, Bloomington, IN 47408, USA}
\author{O.\ L.\ Caballero}
\affiliation{Department of Physics, University of Guelph, Guelph, Ontario N1G 2W1, Canada}
\author{Zidu Lin}
\affiliation{Center for Exploration of Energy and Matter and Department of Physics, Indiana University, Bloomington, IN 47408, USA}
\author{Evan O'Connor}
\affiliation{Department of Physics, North Carolina State University, Raleigh, NC 27695, USA; Hubble  Fellow}
\author{A. Schwenk}
\affiliation{Institut f\"ur Kernphysik, Technische Universit\"at Darmstadt, 64289 Darmstadt, Germany}
\affiliation{ExtreMe Matter Institute EMMI, GSI Helmholtzzentrum f\"ur Schwerionenforschung GmbH, 64291 Darmstadt, Germany}
\affiliation{Max-Planck-Institut f\"ur Kernphysik, Saupfercheckweg 1, 69117 Heidelberg, Germany}

\begin{abstract}
We extend our virial approach to study the neutral-current neutrino
response of nuclear matter at low densities. In the long-wavelength
limit, the virial expansion makes model-independent predictions for
neutrino-nucleon scattering rates and the density $S_V$ and spin $S_A$
responses. We find $S_A$ is significantly reduced from one even at low
densities. We provide a simple fit $S_A^f(n,T,Y_p)$ of the axial
response as a function of density $n$, temperature $T$ and proton
fraction $Y_p$, which can be incorporated into supernova simulations
in a straight forward manner. This fit reproduces our virial results
at low densities and the Burrows and Sawyer random phase approximation
(RPA) model calculations at high densities. Preliminary one
dimensional supernova simulations suggest that the virial reduction in
the axial response may enhance neutrino heating rates in the gain
region during the accretion phase of a core-collapse supernovae.
\end{abstract}


\maketitle 

\section{Introduction}

Neutrinos radiate $99 \%$ of the energy and play a crucial role in
core-collapse supernovae~\cite{SNJanka,SNBurrows,SNMezza}. The
scattering of neutrinos and their transport of energy to the shock
region are sensitive to the physics of low-density nucleonic matter,
which is a complex problem due to bound nuclei and the strong
correlations induced by nuclear forces. A recent three-dimensional
supernova simulation was sensitive to modest changes in
neutral-current neutrino-nucleon interactions and exploded when
strange-quark contributions were included~\cite{Garching3D}. However,
these strange-quark contributions were probbaly taken to be
unrealistically large~\cite{strange}. In this paper, we explore if
similar reductions in neutral-current interactions can arise, not from
strange-quark contributions but, from correlations in low-density
nucleonic matter. The physics of neutrino-matter interactions is a
broad and active field, where many interesting studies of
neutrino-matter interactions have been performed
recently~\cite{Horowitz1991,BS,RPLP,Horowitz2000,Horowitz2004,Burrows2006,Horowitz2012,Bacca2012,Fischer2013,Bartl2014,Rrapaj2015,Balasia2015,Sharma2015,Fischer2016,Bartl2016}.

For low densities and high temperatures, the virial expansion provides
a model-independent approach. In previous works, we have presented the
virial equation of state of low-density nuclear matter~\cite{vEOSnuc}
and of pure neutron matter~\cite{vEOSneut}. In particular, the virial
expansion can be used to describe matter in thermal equilibrium around
the neutrinosphere in supernovae. The temperature of the
neutrinosphere is roughly $T \sim 4$~MeV from about $20$ neutrinos
detected in SN1987a~\cite{sn1987a1,sn1987a2} and the density is $n
\sim 10^{11}-10^{12} \, \text{g}/\text{cm}^{3}$. For pure neutron
matter, the virial expansion in terms of the fugacity $z = e^{\mu/T}$
is valid for
\begin{equation}
n = \frac{2}{\lambda^3} \, z + {\mathcal O}(z^2) \, \lesssim \,
4 \times 10^{11} \, (T/\text{MeV})^{3/2} \, \text{g}/\text{cm}^{3} \,,
\label{zneutmatt}
\end{equation}
where $\lambda = (2\pi/mT)^{1/2}$ denotes the thermal wavelength. A
conservative validity range of the virial equation of state is given
by $z < 1/2$, which gives the limiting density in
Eq.~(\ref{zneutmatt}). Therefore, the virial approach is applicable
for the conditions of the neutrinosphere. Following our virial
equation of state, we have generalized the approach to study
spin-polarized neutron matter and the consistent neutrino response of
neutron matter at low densities~\cite{response}.

In this paper, we use the virial expansion to describe how neutrinos
interact with low-density nuclear matter composed of protons and
neutrons. We neglect alpha particles and other light
nuclei~\cite{mass3,light}. These will be included in later work. In
Sec.~\ref{response}, we present our formalism. Our results for the
axial response and preliminary one-dimensional supernova simulations
are presented in Sec.~\ref{results}. Finally, we conclude in
Sec.~\ref{conclusions}.

\section{Neutrino response}
\label{response}

In this section, we use the virial expansion to describe how neutrinos
interact with low-density nuclear matter.  We focus on neutral-current
neutrino interactions.  We expect similar results for charged-current
reactions, however we leave these to later work.  We calculate the
neutrino cross section per unit volume.  The virial expansion provides
model-independent results in the limit of low momentum transfer $q
\rightarrow 0$.

The free cross section for neutrino-nucleon neutral-current scattering is
\begin{equation}
\frac{d\sigma_0}{d\Omega}_{\nu N}=\frac{G_F^2 E_\nu^2}{4\pi^2} \Bigl({C_a^N}^2
(3-\cos\theta)+{C_v^N}^2 (1+\cos\theta) \Bigr) \,,
\label{sigmanuN}
\end{equation}
where $G_F$ is the Fermi constant, $E_\nu$ the neutrino energy, and
$\theta$ the scattering angle. The axial coupling up to strange-quark
corrections is $|C_a^N| = |g_a|/2 = 0.63$ where $g_a$ is the axial
charge of the nucleon. The weak vector charge is $C_v^n=-1/2$ for
scattering from a neutron $n$ and $C_v^p=1/2-2\sin^2\theta_W \approx
0$ for scattering from a proton $p$. Here $\theta_W$ is the weak
mixing angle. The cross section in Eq.~(\ref{sigmanuN}) neglects
corrections of order $E_\nu/m$ from weak magnetism and other effects,
for details see~\cite{weakmag}.

The free cross section per unit volume for scattering from a mixture of
neutrons and protons is then given by
\begin{align}
\frac{1}{V} \frac{d\sigma_0}{d\Omega} &= n_n \, \frac{d\sigma_0}{d\Omega}_{\nu n}+ n_p \, \frac{d\sigma_0}{d\Omega}_{\nu p} \,, \\[2mm]
&= \frac{G_F^2 E_\nu^2}{16\pi^2} \Bigl( g_a^2 (3-\cos\theta)(n_n+n_p) \nonumber \\
& \quad\quad\quad\quad + (1+\cos\theta)n_n \Bigr) \,. \label{sigma0}
\end{align}
In the medium this cross section is modified by the density (vector)
$S_V$ and the spind (axial) $S_A$ response.  The response of the
system to density fluctuations is described by $S_V$, while $S_A$
describes the response of the system to spin fluctuations.  The
response functions are normalized to unity in the low-density limit
$S_V, S_A \rightarrow 1$ as $n \rightarrow 0$.  The cross section per
unit volume in the medium is then given by
\begin{align}
\frac{1}{V} \frac{d\sigma}{d\Omega} &= \frac{G_F^2 E_\nu^2}{16\pi^2} 
\Bigl(g_a^2(3-\cos\theta)(n_n+n_p)S_A \nonumber \\ 
& \quad\quad\quad\quad +(1+\cos\theta)n_n S_V \Bigr) \,.
\label{sigma}
\end{align}
Note that $d\sigma/d\Omega$ reduces to the free cross section
$d\sigma_0/d\Omega$ as $S_A, S_V \rightarrow 1$.  In general both
$S_V$ and $S_A$ depend on momentum transfer $q$.  However, in the
limit $q \rightarrow 0$ we can derive model independent virial
results.

\subsection{Virial equation of state}

Next, we briefly review the virial equation of state for a system with
neutrons and protons~\cite{vEOSnuc}.  We will use this to calculate
$S_V$ and $S_A$.  The pressure $P$ is expanded to second order in the
fugacities of neutrons $z_n$ and protons $z_p$,
\begin{equation}
\frac{P}{T} = \frac{\ln Q}{V}= \frac{2}{\lambda^3}\Bigl[z_n+z_p+(z_n^2+z_p^2)b_n+2z_pz_nb_{pn}\Bigr] \,.
\end{equation}
Here $T$ is the temperature, $V$ is the volume of the system, and $Q$
is the grand-canonical partition function.  The fugacities are related
to the neutron $\mu_n$ and proton $\mu_p$ chemical potentials by
$z_n=e^{\mu_n/T}$ and $z_p=e^{\mu_p/T}$.  Finally the second virial
coefficients $b_n$ and $b_{pn}$ are calculated from nucleon-nucleon
elastic scattering phase shifts.  These are tabulated in
Ref.~\cite{vEOSnuc}.

The neutron $n_n$ and proton $n_p$ densities follow from derivatives
of $\ln Q$,
\begin{equation}
n_i=z_i \frac{\partial}{\partial z_i}\Bigl(\frac{\ln Q}{V}\Bigr)\Bigr|_{V,T}\, .
\end{equation}
This gives
\begin{align}
n_n &= \frac{2}{\lambda^3}(z_n+2z_n^2b_n+2z_pz_nb_{pn}) \,,
\label{equ:ndens} \\[1mm]
n_p &= \frac{2}{\lambda^3}(z_p+2z_p^2b_n+2z_pz_nb_{pn}) \,.
\end{align}

\subsection{Vector response}

\begin{figure}[t]
\centering
\includegraphics[width=\columnwidth]{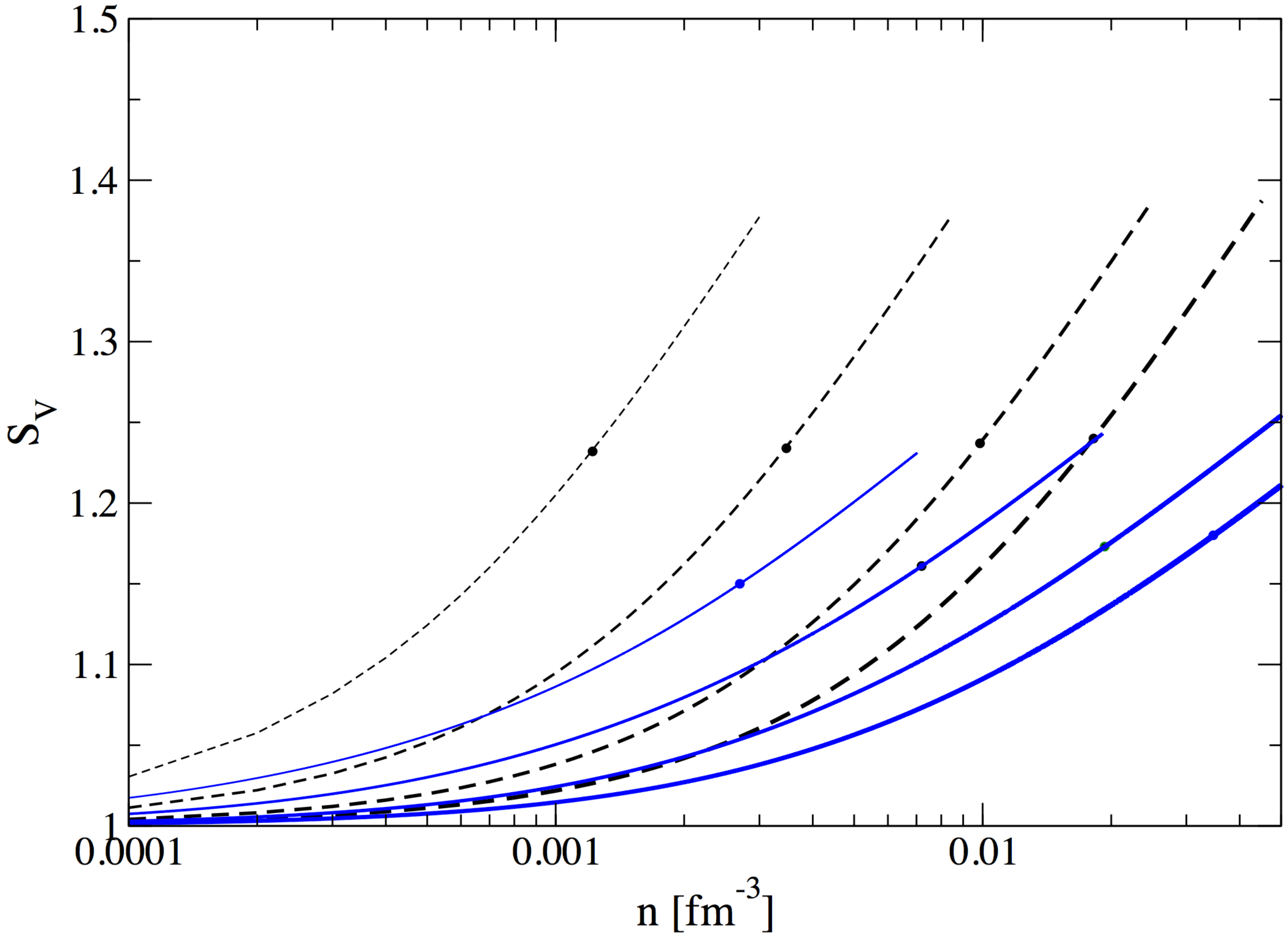}
\caption{(Color online) Vector response $S_V$ versus density $n$ for 
proton fractions of $Y_p=0$ dashed and $0.3$ solid lines. The curves
are for temperatures of (left to right with increasing thickness) 2.5,
5, 10, and 15~MeV. The solid circles show where $z_n=0.5$. The virial
expansion is most valid to the left of these points.}
\label{fig.Sv}
\end{figure}

The vector response $S_V$ is equal to the static structure factor
$S_q$, see for example, Refs.~\cite{LL,response}.  For a
single-component system
\begin{equation}
S_V(q=0) = \frac{T}{(\partial P/\partial n)_T} \,.
\end{equation}
Using the virial equation of state this can be rewritten with
$dP/dn=n/(Tz)(dz/dn)$ as
\begin{equation}
S_V=\frac {1}{n} z \frac{\partial}{\partial z} n \,.
\end{equation}
Following Ref.~\cite{BS}, we generalize this result to a mixture of
neutrons and protons:
\begin{equation}
S_V = \frac{{C_v^n}^2 S_{nn} + 2 C_v^n C_v^p S_{np} + {C_v^p}^2 S_{pp}}{{C_v^n}^2 n_n + {C_v^p}^2 n_p} \,,
\end{equation}
where
\begin{align}
S_{nn} &= z_n \frac{\partial}{\partial z_n} n_n = n_n+\frac{4}{\lambda^3}z_n^2 b_n \,, \label{nn} \\[2mm]
S_{np} &= z_p \frac{\partial}{\partial z_p} n_n = \frac{4}{\lambda^3}z_pz_n b_{pn} \,, \label{np} \\[2mm]
S_{pp} &= z_p \frac{\partial}{\partial z_p} n_p = n_p+\frac{4}{\lambda^3}z_p^2 b_n \,. \label{pp}
\end{align}
Using Eqs.~(\ref{nn},\ref{np},\ref{pp}), we have for $S_V$
\begin{equation}
S_V = 1 + \frac{4}{\lambda^3} \, \frac{{C_v^n}^2z_n^2b_n+2C_v^nC_v^p z_nz_p b_{pn} + {C_v^p}^2z_p^2 b_n}{{C_v^n}^2n_n+{C_v^p}^2n_p} \,.
\label{svfull}
\end{equation}
In the limit $C_v^p \approx 0$ this reduces to the neutron-matter
result~\cite{response}
\begin{equation}
S_V = 1 + \frac{4}{\lambda^3} \, \frac{z_n^2 b_n}{n_n} \,.
\label{Svfinal}
\end{equation}
Here the impact of protons is to somewhat modify the neutron fugacity
$z_n$ because of the $b_{pn}$ term in the neutron density,
Eq.~(\ref{equ:ndens}).  The virial coefficient $b_n \approx 0.32$ is
small and positive.  As a result the vector response is slightly
enhanced (larger than one) as shown in Fig.~\ref{fig.Sv}.  Attractive
nucleon-nucleon interactions increase the probability to find nucleons
close together.  These density fluctuations increase the (local) weak
charge and produce a vector response $S_V>1$.

\subsection{Axial response}

To calculate the axial response $S_A$ we generalize our virial
equation of state to describe spin-polarized nuclear matter.  Let
$z_p^+$, $z_n^+$ be the fugacities for spin up $p$ and $n$, and
$z_p^-$, $z_n^-$ be the spin down fugacities.  Generalizing the
results of Ref.~\cite{response}, we have for the density of spin-up
neutrons $n_n^+$
\begin{equation}
n_n^+=\frac{1}{\lambda^3} \Bigl[ z_n^+ + 2 b_+ {z_n^+}^2 + 2 z_n^+(b_-z_n^- + b_{pn}^+ z_p^++ b_{pn}^- z_p^-)\Bigr] \,.
\end{equation}
We discuss the spin virial coefficients $b_+$, $b_-$, $b_{pn}^+$ and
$b_{pn}^-$ in Sec. \ref{sec:spinvirial}.  Likewise the density of spin-down neutrons $n_n^-$ is
\begin{equation}
n_n^-=\frac{1}{\lambda^3}\Bigl[ z_n^- + 2 b_+ {z_n^-}^2 + 2 z_n^-(b_-z_n^+ + b_{pn}^+ z_p^-  + b_{pn}^- z_p^+) \Bigr] \,.
\end{equation}
Similarly, the density of spin-up protons $n_p^+$ is
\begin{equation}
n_p^+=\frac{1}{\lambda^3}\Bigl[z_p^+ + 2 b_+ {z_p^+}^2 + 2 z_p^+(b_-z_p^- + b_{pn}^+ z_n^+ + b_{pn}^- z_n^-) \Bigr] \,,
\end{equation}
while the density of spin-down protons $n_p^-$ is
\begin{equation}
n_p^-=\frac{1}{\lambda^3}\Bigl[z_p^- + 2 b_+ {z_p^-}^2 + 2 z_p^-(b_-z_p^+ + b_{pn}^+ z_n^- + b_{pn}^- z_n^+) \Bigr] \,.
\end{equation}
We define axial or spin fugacities $z_n^a=(z_n^+/z_n^-)^{1/2}$ and
$z_p^a=(z_p^+/z_p^-)^{1/2}$ and following Ref.~\cite{BS} write the
axial response as
\begin{equation}
S_A=\frac{S_{pp}^A+S_{nn}^A-2S_{np}^A}{n_n+n_p} \,,
\label{sa}
\end{equation}
with
\begin{align}
S_{nn}^A &= z_n^a \frac{\partial}{\partial z_n^a} (n_n^+-n_n^-)\Bigl|_{z_n^a=1}
=n_n+\frac{4}{\lambda^3}(b_+ - b_-)z_n^2 \,, \\[2mm]
S_{np}^A &= z_p^a \frac{\partial}{\partial z_p^a} (n_n^+-n_n^-)\Bigl|_{z_p^a=1}
=\frac{4}{\lambda^3}z_nz_p(b_{pn}^+-b_{pn}^-) \,, \\[2mm]
S_{pp}^A &= z_p^a \frac{\partial}{\partial z_p^a} (n_p^+-n_p^-)\Bigl|_{z_p^a=1}
=n_p+\frac{4}{\lambda^3}(b_+ - b_-)z_p^2 \,.
\end{align}
Note that the minus sign for the $S_{np}^A$ term in Eq.~(\ref{sa}) is
because the axial charge of a neutron is opposite to that of a
proton. To clean up the notation, we define the axial virial
coefficients
\begin{align}
b_a &= b_+-b_- \,, \\
b_{pn}^a &= b_{pn}^+-b_{pn}^- \,.
\end{align}
The final result for $S_A$ can then be written as
\begin{equation}
S_A = 1+\frac{4}{\lambda^3} \, \frac{(z_p^2+z_n^2)b_a-2z_pz_nb_{pn}^a}{n_n+n_p} \,.
\label{safinal}
\end{equation}   
To lowest order in the density one has $z_p \approx \lambda^3 n_p/2$
and $z_n \approx \lambda^3 n_n/2$ so that
\begin{equation}
S_A \approx 1+ \lambda^3 \, \frac{(n_n^2+n_p^2)b_a-2n_nn_pb^a_{pn}}{n_p+n_n} \,.
\label{saapprox}
\end{equation}
Note that we use the full Eq.~(\ref{safinal}) for results in the next
section.  Because the spin virial coefficient $b_a \approx -0.6$ (see
below), the axial response is reduced $S_A<1$.  This is because two
neutrons or two protons are likely to be correlated in a $^1$S$_0$
state because of the Pauli principle and this spin zero state reduces
the spin response.

We define the total response $S_{\rm tot}$ as the ratio of the
in-medium transport cross section to the free one,
\begin{equation}
S_{\rm tot}=\frac{\int d\Omega \, \frac{d\sigma}{d\Omega} (1-\cos\theta)}{\int d\Omega \, \frac{d\sigma_0}{d\Omega} (1-\cos\theta)} \,.
\end{equation}
From Eqs.~(\ref{sigma0}) and~(\ref{sigma}), we thus have
\begin{equation}
S_{\rm tot} = \frac{5g_a^2 S_A + (1-Y_p) S_V}{5g_a^2 + 1-Y_p} \,,
\label{eq.stot}
\end{equation}
where $Y_p$ is the proton fraction. The total response depends on both
$S_V$ and $S_A$.  However, in general $S_A$ is most important because
of the large factor $5g_a^2$.  We present results for $S_A$ in
Sec.~\ref{results}.  However first we discuss the spin virial
coefficients.

\subsection{Spin virial coefficients}
\label{sec:spinvirial}

The virial coefficient $b_a=b_+-b_-$ is discussed in Ref. \cite{response} and
describes spin interactions between two protons or two neutrons.  We
now discuss $b_{pn}^a=b_{pn}^+-b_{pn}^-$ that involves interactions
between protons and neutrons.  The virial coefficient $b_{pn}^+$
describes interactions between a $p$ and a $n$ with like spin
projections, while $b_{pn}^-$ describes interactions between nucleons
with unlike spins. We therefore have
\begin{multline}
b_{pn}^a = \frac{1}{2^{1/2}}(e^{E_d/T}-1) \\ + \frac{2^{1/2}}{\pi T} \int_0^\infty dE e^{-E/2T} [\delta_{pn}^+(E)-\delta_{pn}^-(E)] \,,
\end{multline}
with
\begin{multline}
\delta_{pn}^+(E) = \frac{1}{2}\delta_{^3S_1}+\frac{1}{6}\delta_{^3P_0}+\frac{1}{2}\delta_{^3P_1}+\frac{5}{6}\delta_{^3P_2} \\
+\frac{1}{2}\delta_{^3D_1}+\frac{5}{6}\delta_{^3D_2}+\frac{7}{6}\delta_{^3D_3} + \ldots \,.
\end{multline}
Here $E_d$ is the deuteron binding energy and the factor in front of
each phase shift is $(2J+1)/[2(2S+1)]$ where the factor of $1/2$ is
from the average over isospin 1 and 0 states. In our calculation, we
have neglected states with $L>2$. Similarly, we have
\begin{multline}
\delta_{pn}^-(E)=\frac{1}{4}\delta_{^1S_0}+\frac{1}{4}\delta_{^3S_1}+\frac{3}{4}\delta_{^1P_1}+\frac{1}{12}\delta_{^3P_0}+\frac{1}{4}\delta_{^3P_1} \\
+\frac{5}{12}\delta_{^3P_2}+\frac{5}{4}\delta_{^1D_2}+\frac{1}{4}\delta_{^3D_1}+\frac{5}{12}\delta_{^3D_2}+\frac{7}{12}\delta_{^3D_3} + \ldots \,.
\end{multline}
Now the factor for each phase shift is $(2J+1)/[4(2S+1)]$, where the
$1/4$ is from an average over both isospin 1 and 0 and spin 1 and 0
states. We calculate the spin virial coefficients based on the
Nijmegen partial-wave analysis of nucleon-nucleon
scattering~\cite{nnphases}. Our results for $b_a$ and $b_{pn}^a$ are
collected in Table~\ref{table1}.  The other virial coefficients $b_n$
and $b_{pn}$ needed to calculate the neutrino responses have all ready
been tabulated in Ref.~\cite{vEOSnuc}.

In contrast to $b_a$, $b_{pn}^a$ is positive because a proton and a
neutron can be correlated into the spin one $^3$S$_1$ state (deuteron
like), enhancing the spin response.  However the axial charge of a
proton is opposite to that of a neutron.  This leads to a minus sign
in Eq.~(\ref{safinal}) for the $b_{pn}^a$ term.  As a result, both
$b_a$ and $b_{pn}^a$ reduce the total axial response and lead to
$S_A<1$.

\begin{table}
\begin{tabular}{l|l|l}
$T$~(MeV) \: & \: $b_a$ \: & \: $b_{pn}^a$ \\
\hline
1\: & \: -0.638\: & \: 6.18 \\
2\: & \: -0.653\: & \: 1.74 \\
3\: & \: -0.651\: & \: 1.05 \\
4\: & \: -0.648\: & \: 0.785 \\
5\: & \: -0.643\: & \: 0.640 \\
6\: & \: -0.637\: & \: 0.561 \\
7\: & \: -0.631\: & \: 0.504 \\
8\: & \: -0.625\: & \: 0.463 \\
9\: & \: -0.620\: & \: 0.432 \\
10\: & \: -0.615\: & \: 0.408 \\
12\: & \: -0.605\: & \: 0.374 \\
14\: & \: -0.597\: & \: 0.352 \\
16\: & \: -0.589\: & \: 0.336 \\
18\: & \: -0.583\: & \: 0.324 \\
20\: & \: -0.577\: & \: 0.315
\end{tabular} 
\caption{Spin virial coefficients $b_a$ and $b_{pn}^a$ based on 
nucleon-nucleon phase shifts.} 
\label{table1}
\end{table}

\begin{figure*}[t]
\begin{tabular}{cc}
\includegraphics[width=\columnwidth]{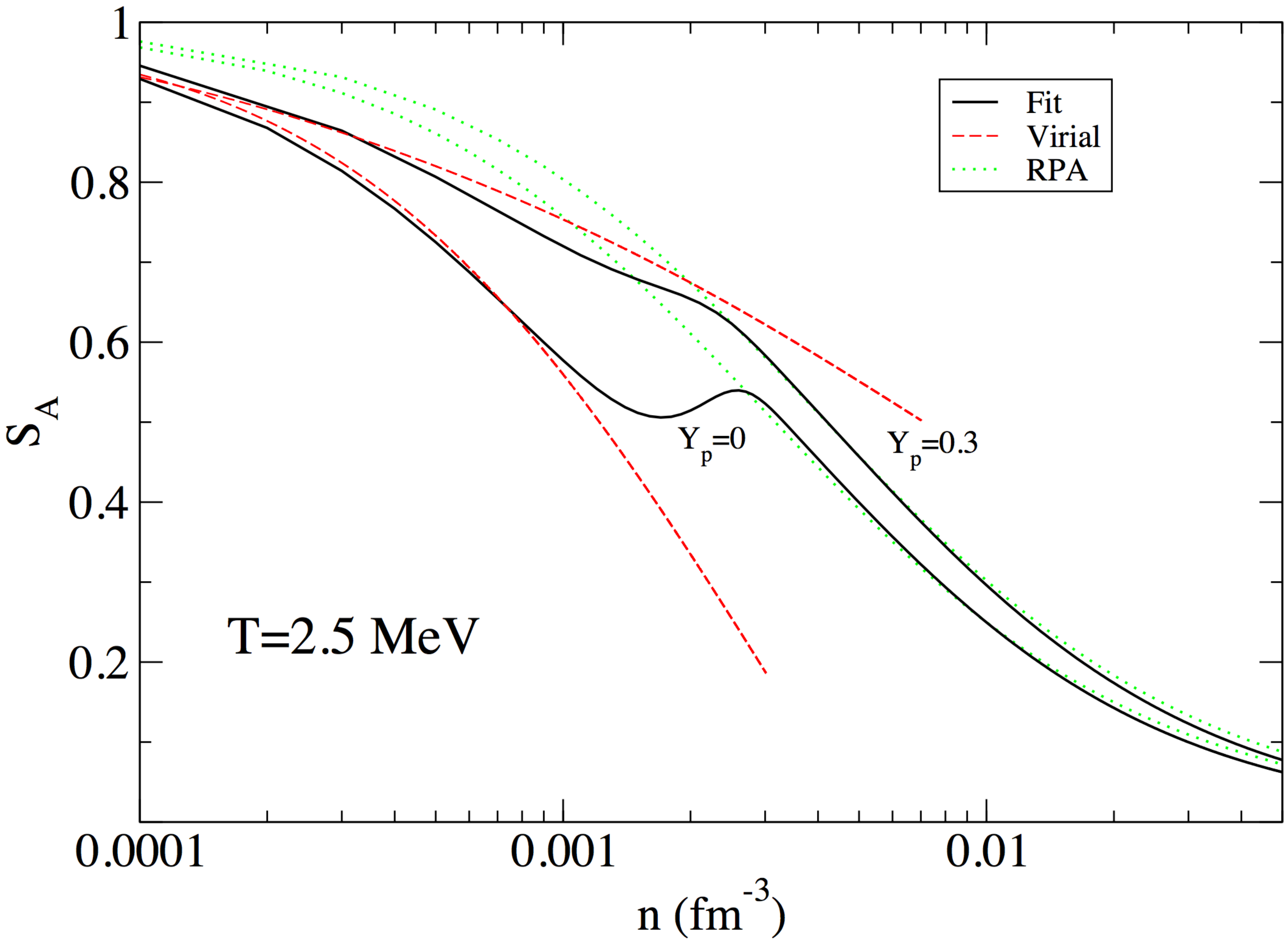} \: & \: 
\includegraphics[width=\columnwidth]{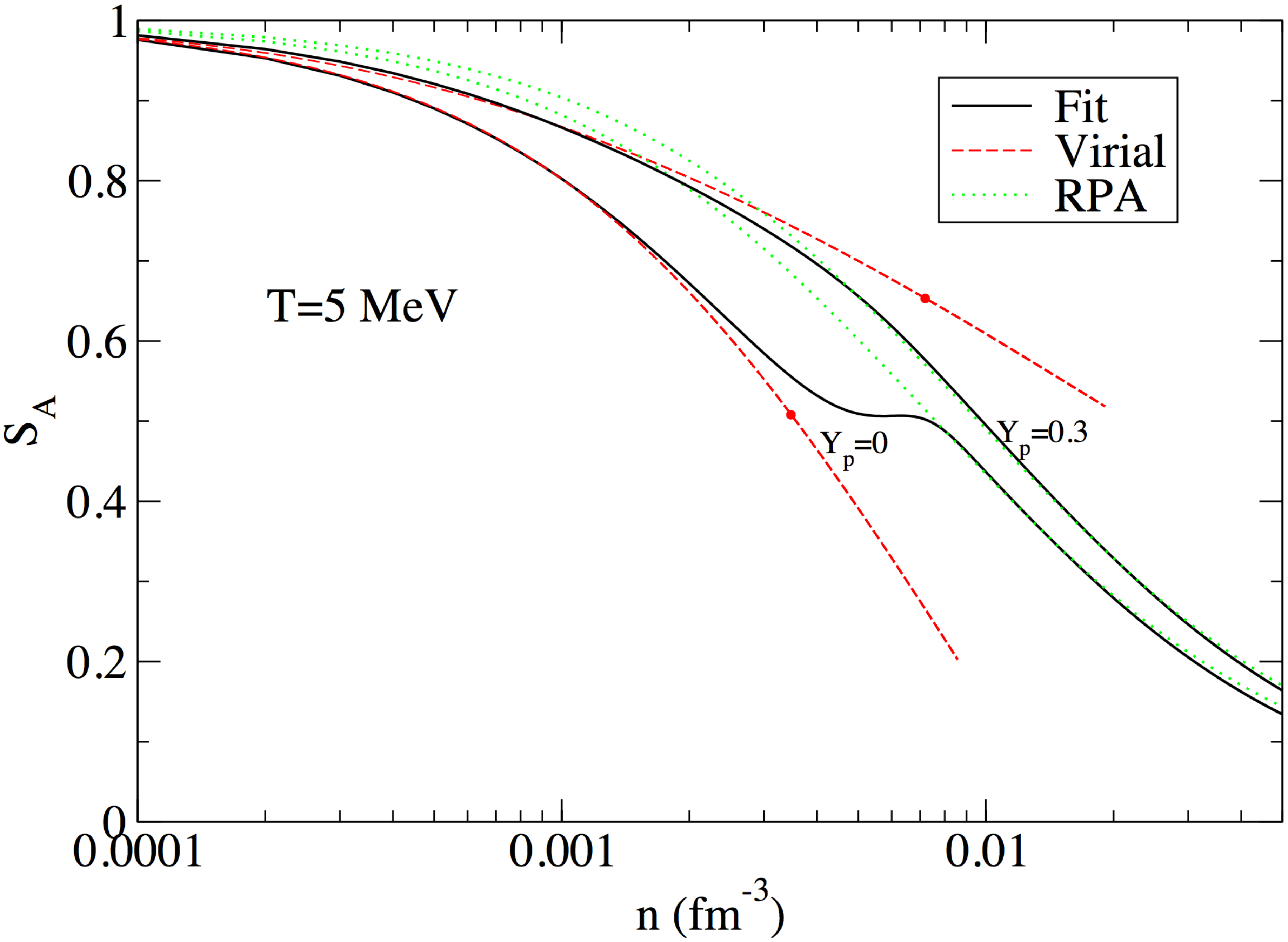} \\ 
(a) & (b) \\[2mm]
\includegraphics[width=\columnwidth]{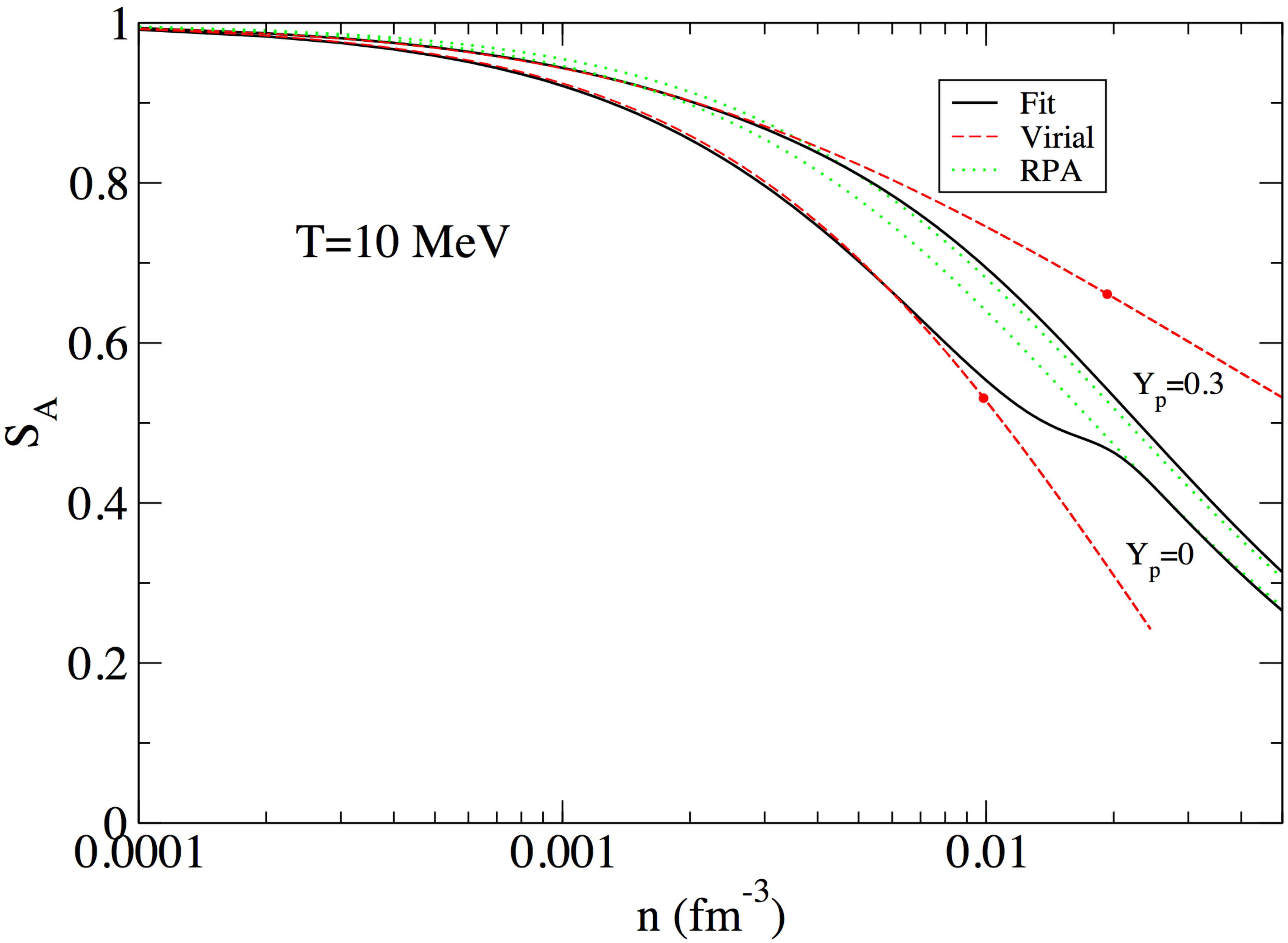} \: & \: 
\includegraphics[width=\columnwidth]{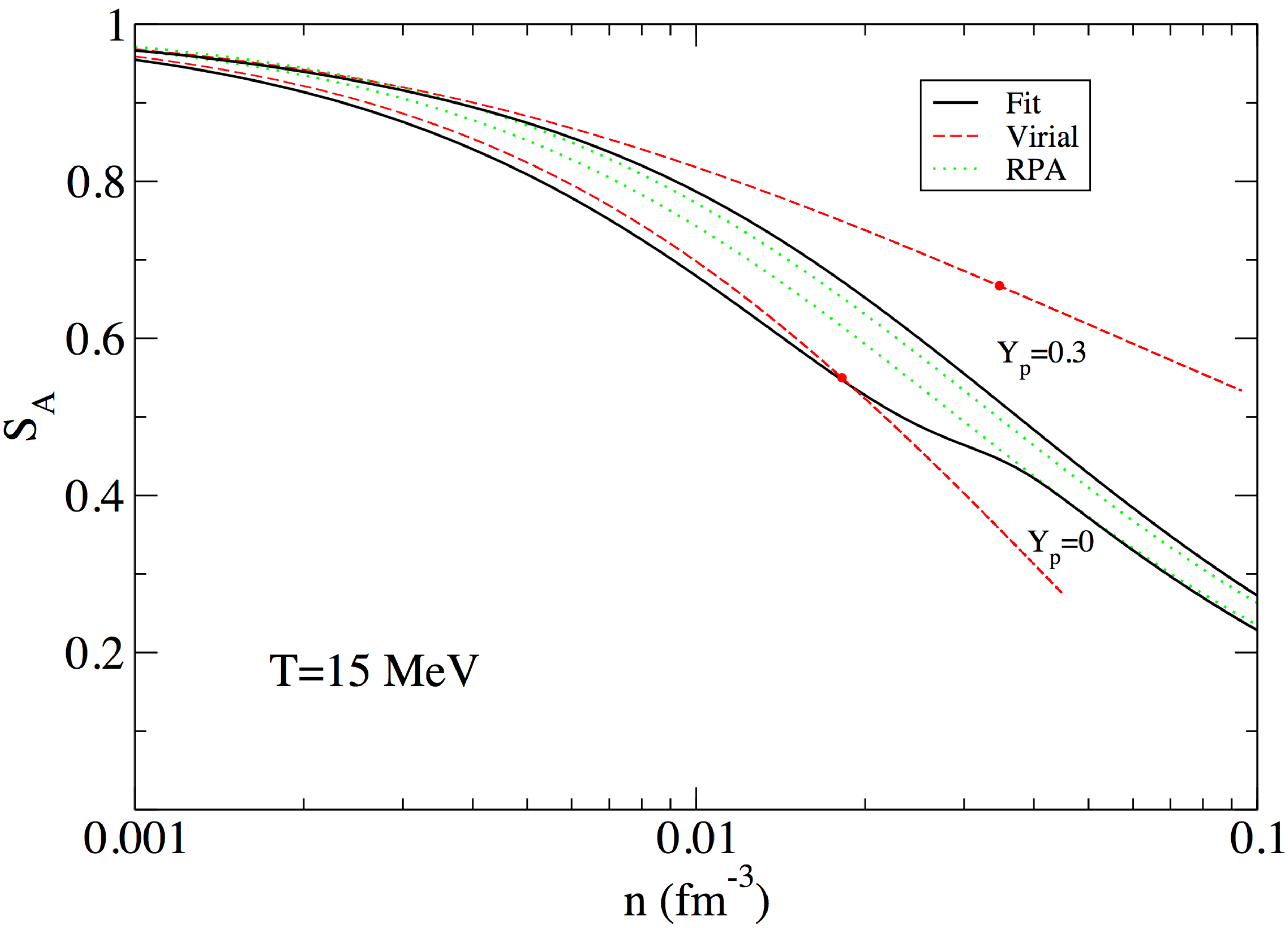} \\ 
(c) & (d)
\end{tabular}
\caption{(Color online) Axial response $S_A$ versus density $n$ for 
temperatures of $T=2.5$~MeV (a), 5~MeV (b), 10~MeV (c), and 15~MeV
(d).  The red dashed lines are our virial expansion results,
Eq.~(\ref{safinal}), for the indicated proton fractions.  The solid
red dots indicate where $z_n=0.5$.  The virial expansion is most valid
to the left of these points.  The green dotted lines show the original
Burrows and Sawyer RPA results~\cite{BS}.  Finally the solid black
lines show the interpolating fit $S_A^{f}$, Eq.~(\ref{eq.Saf}).}
\label{fig.Sa}
\end{figure*}

\subsection{Combining correlations, strange-quark, weak magnetism, and recoil corrections}

We end this formalism section by describing the combination of
correlation, strange-quark, weak magnetism, and recoil corrections.
To a good approximation all of these effects can be combined in a
straight forward way that avoids double counting. We write for the
neutrino cross section per unit volume, see Eq.~(\ref{sigma}),
\begin{multline}
\frac{1}{V}\frac{d\sigma}{d\Omega} \approx \frac{G_F^2 E_\nu^2}{16 \pi^2}
\Bigl( \bigl[(g_a+g_a^s)^2n_n+(g_a-g_a^s)^2n_p \bigr] \\
(3-\cos\theta)S_A+(1+\cos\theta)n_nS_V\Bigr) R(E_\nu/m) \,.
\label{sigmatot}
\end{multline}
Effects from nucleon-nucleon correlations in the medium are described
by the vector $S_V$ and axial $S_A$ response functions, see
Eqs.~(\ref{Svfinal}) and~(\ref{safinal}), respectively.  The vector
response is slightly greater than one and the axial response is
significantly less than one.  As a result correlations reduce the
cross section for both neutrino-proton and neutrino-neutron
scattering.

Strange quark contributions to the nucleon spin are described by the
parameter $g_a^s$ (note $g_a=1.26$).  Melson {\it et
al.}~\cite{Garching3D} consider $g_a^s=-0.2$ and this value reduces
neutrino-neutron and increases neutrino-proton scattering cross
sections. For neutron-rich conditions this leads to a net reduction in
the neutrino scattering opacity. Therefore both correlation effects
and strange quarks, if present (with $g_a^s<0$), reduce the opacity,
and the two effects add. Note that in Ref.~\cite{Garching3D}
strange-quark contributions were probbaly taken to be unrealistically
large~\cite{strange}. On the one side, strange-quark contributions to
the vector current have been measured by several parity-violating
electron scattering experiments to be small~\cite{PV}, but direct
experimental limits on strange-quark contributions to the nucleon spin
are relatively poor and are based on an old Brookhaven neutrino
scattering experiment~\cite{Brookhaven}.  Therefore, it would be very
useful to have a better laboratory limit on $g_a^s$ from a modern
neutrino-nucleon scattering experiment.

Finally recoil and weak magnetism corrections can be approximately
described by a factor $R(E_\nu/m)$. This is discussed in
Ref.~\cite{weakmag} and reduces antineutrino-nucleon scattering cross
sections, while having only a modest effect on neutrino-nucleon cross
sections.

\section{Results for the Axial Response}
\label{results}

In this section, we focus on results for the axial response $S_A$, and
not on the vector response $S_V$, for two reasons.  First, the axial
response is more important for neutrino-transport cross sections
because of a large $5g_a^2$ factor, see Eq.~(\ref{sigmatot}).  Second,
we have not included alpha particles or other light nuclei.
Preliminary calculations suggest that spin zero alpha particles can
significantly enhance $S_V$, but do not strongly impact $S_A$.
Therefore we postpone a full discussion of $S_V$ to later work, where
we will explicitly include alpha particles and other light nuclei.
For the present, a reasonable approximation is to simply set $S_V = 1$
in Eq.~(\ref{eq.stot}).

In Fig.~\ref{fig.Sa} we show $S_A$ for temperatures of $T=2.5$ to
15~MeV.  Our virial results (red dashed lines) are valid at low
densities.  To evaluate $S_A$ for higher densities, where $z_n>0.5$,
one presently needs to employ a model-dependent calculation.  We
consider the random phase approximation (RPA) calculations of Burrows
and Sawyer~\cite{BS}, because they are simple, well known, and have
been employed in supernova simulations.  We caution that these
calculations may have a number of limitations.  First they predict
that the vector response is less than one $S_V<1$ while
Fig. \ref{fig.Sv} shows $S_V>1$.  Second the calculations use a Landau
parameter for the effective interaction that is appropriate for
symmetric nuclear matter.  A Landau parameter appropriate for pure
neutron matter could lead to a smaller $S_A$. See also the discussion
in Ref.~\cite{response}. In future work we will revisit the behavior
of $S_A$ at high densities, but for now we consider the Burrows and
Sawyer results.  The green dotted lines in Fig.~\ref{fig.Sa} show the
Burrows and Sawyer RPA calculation~\cite{BS}.  Note that the RPA
depends very weakly on the momentum transfer $q$ and we use $q=3T$.

\begin{table}
\begin{tabular}{l|l}
Fit parameter \: & \: Value \\
\hline
$A_0$ & \: 920 \\
$B_0$ & \: 3.05\\
$C_0$ & \: 6140 \\
$D_0$ & \: $1.5 \times 10^{13}$\\
\end{tabular} 
\caption{Fit parameters for $S_A^f$ fitting function, see 
Eqs.~(\ref{eq.Saf}--\ref{eq.Safc}). 
These assume $n$ in fm$^{-3}$ and $T$ in MeV.} 
\label{table2}
\end{table}

Finally, we fit the virial results for $S_A$ at low densities and the
RPA results at high densities with an interpolating function
$S_A^f(n,T,Y_p)$ that is a simple function of density $n$, temperature
$T$, and proton fraction $Y_p$:
\begin{equation}
S_A^f(n,T,Y_p) = \frac{1}{1+A (1+B e^{-C})} \,,
\label{eq.Saf}
\end{equation}
where the functions $A$, $B$, and $C$ are given by
\begin{align}
A(n,T,Y_p) &= A_0\frac{n(1-Y_p+Y_p^2)}{T^{1.22}} \,,
\label{eq.Safa} \\
B(T) &= \frac{B_0}{T^{0.75}} \,,
\label{eq.Safb} \\
C(n,T,Y_p) &= C_0 \frac{nY_p(1-Y_p)}{T^{0.5}}+D_0\frac{n^4}{T^6} \,.
\label{eq.Safc}
\end{align}
The fit parameters $A_0$, $B_0$, $C_0$, and $D_0$ are collected in
Table~\ref{table2} for $n$ in fm$^{-3}$ and $T$ in MeV.  This fit is
most accurate for $5<T<10$ MeV, $Y_p\le 0.3$, and $n<0.05$ fm$^{-3}$,
but yields reasonable values outside this range.  We make an
implementation of this fitting function available in
NuLib~\cite{oconnor:15} at \url{http://www.nulib.org}.  Note that if
one sets $B_0=0$ in Eq.~(\ref{eq.Safb}), Eq.~(\ref{eq.Saf}) will
approximately reproduce the original Burrows and Sawyer RPA results at
low density.  We see that $S_A$ is significantly reduced from 1 even
at relatively low densities.  This is especially the case at low $Y_p$.

\begin{figure}[t]
\centering
\includegraphics[width=\columnwidth]{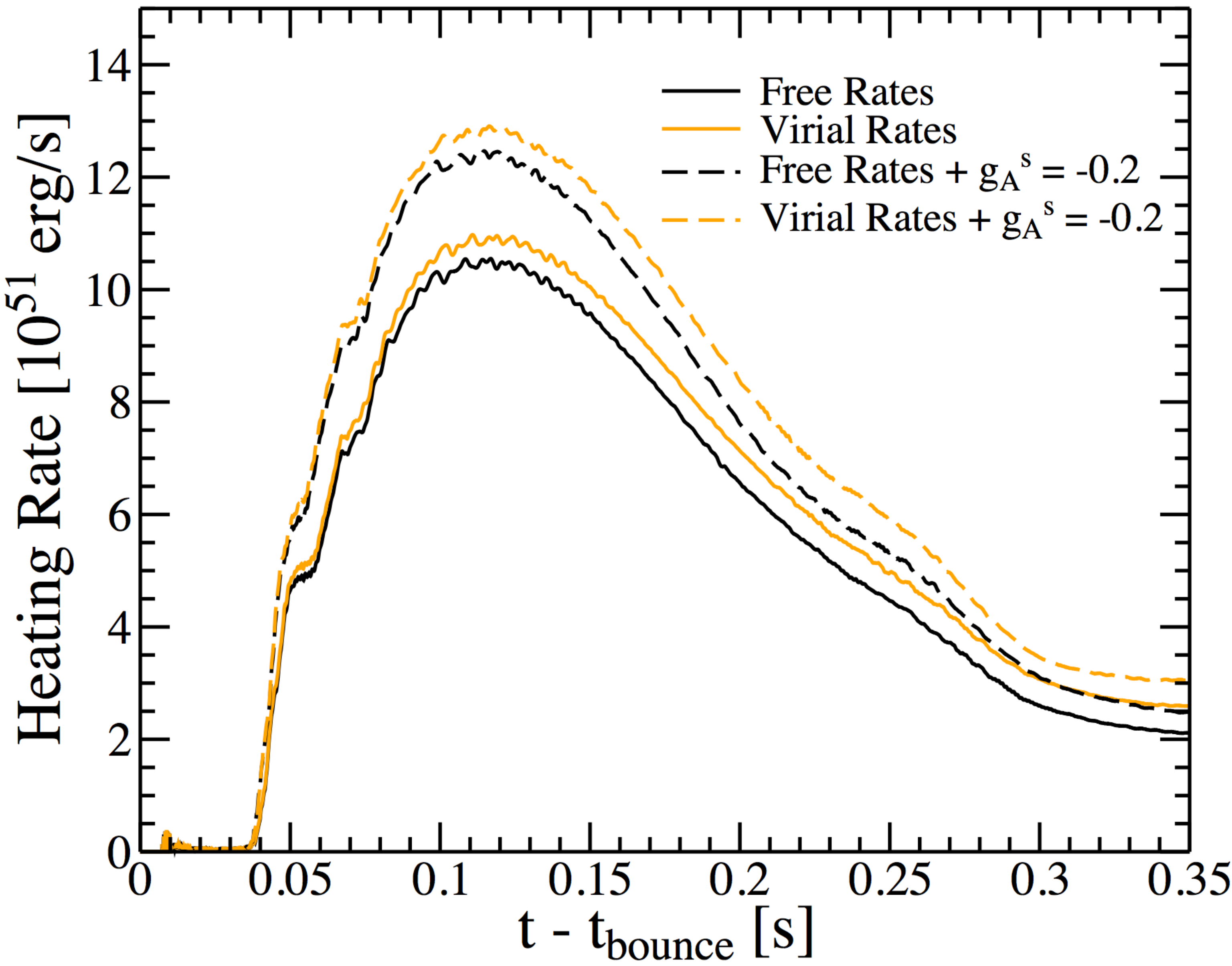}
\caption{(Color online) Heating rate in the gain region for
one-dimensional simulations of the accretion phase of core-collapse
supernovae for various assumptions on the neutral-current
neutrino-nucleon scattering cross section. The solid lines represent
simulations that ignore the contribution to the rates from strange
quarks, whereas dashed lines denote simulations that include
strange-quark contributions, assuming $g_a^s=-0.2$ for comparison with
Ref.~\cite{Garching3D}. The black lines show the heating rate for
simulations that assume $S_A=1$, while orange lines show the larger
heating rates with $S_A^f$ from Eq.~(\ref{eq.Saf}).}
\label{fig:heatingrates}
\end{figure}

We briefly explore the effect of the reduced axial response arising
from the virial expansion on the heating rate obtained in simulations
of the accretion phase of a core-collapse supernovae. To compare with
Ref.~\cite{Garching3D}, we use the Lattimer and Swesty \cite{lseos:91}
equation of state (with $K_0=220$~MeV) and the 20\,$M_\odot$
progenitor star from Ref.\cite{woosley:07}.  We use the one
dimensional code GR1D~\citep{oconnor:15}, available at
\url{http://www.GR1Dcode.org}. For the sake of comparison with
Ref.~\cite{Garching3D}, we perform simulations with and without
strange-quark contributions. We show the heating rate realized in
the gain region in Fig.~\ref{fig:heatingrates}. The solid lines denote
simulations using no strange quark corrections, while the dashed lines
denote simulations using $g_a^s =-0.2$. The black lines are for
simulations that use free particle rates, while the orange lines
denote simulations using the virial corrected rates. Similar to
Ref.~\cite{Garching3D}, the assumed strange-quark contribution raises
the heating rate by $\sim 20\%$ around 100\,ms after bounce. We find
the virial rates increase the heating by $\sim 5\%$ around 100\,ms
after bounce and higher at later times.  We expect the reduction in
$S_A$ to play an even larger role at later times when the neutrinos
decouple from regions with higher matter densities.  
    
\section{Summary and Conclusions}
\label{conclusions}

Supernova simulations may be sensitive to the neutral-current
interactions of mu and tau neutrinos at low densities near the
neutrinosphere.  In this paper, we have calculated the axial or spin
response $S_A$ of nuclear matter in a virial expansion that is
model-independent at low densities and high temperatures.  We find
$S_A$ to be significantly reduced.  Our results can be incorporated
into supernova simulations by multiplying both the neutrino-proton and
neutrino-neutron neutral current scattering rates by $S_{\rm tot}$ given
by Eq.~(\ref{eq.stot}) with $S_V=1$ and $S_A$ given by our fit
function $S_A^f$ from Eqs.~(\ref{eq.Saf}), (\ref{eq.Safa}),
(\ref{eq.Safb}), and~(\ref{eq.Safc}).  Preliminary one-dimensional
supernova simulations suggest that the reduction in the axial response
may enhance the neutrino heating rates in the gain region during the
accretion phase of a core-collapse supernova.  In future work, we will
extend the calculation to the vector response $S_V$ in a virial
expansion and study the impact of light nuclei.

\begin{acknowledgments}

We thank Adam Burrows and Thomas Janka for helpful discussions. This
work was supported in part by DOE Grants DE-FG02-87ER40365 (Indiana
University) and DE-SC0008808 (NUCLEI SciDAC Collaboration), by the
Natural Sciences and Engineering Research Council of Canada (NSERC),
and the Deutsche Forschungsgemeinschaft Grant SFB 1245. Support for
this work was provided also by NASA through Hubble Fellowship Grant
\#51344.001-A awarded by the Space Telescope Science Institute, which
is operated by the Association of Universities for Research in
Astronomy, Inc., for NASA, under contract NAS 5-26555.

\end{acknowledgments}

\end{document}